\documentstyle[aps,prbbib,graphicx,floats,amssymb,times]{revtex}
\newcommand{\ek}{\epsilon_{\mathbf{k}}}

\newcommand{\Ek}{E_{\mathbf{k}}}

\newcommand{\phik}{\varphi_{\mathbf{k}}}

\newcommand{\mb}[1]{{\mathbf{#1}}}
\newcommand{\sumk}{\sum_{\mathbf{k}}}

\begin{document}
\draft

\wideabs{

  \title{Superconducting phase coherence in the presence of a pseudogap:
    Relation to specific heat, tunneling and vortex core spectroscopies}
  
  \author{Qijin Chen\cite{byline} and K. Levin} 
  \address{James Franck Institute, University of Chicago, Chicago, 
   Illinois 60637}

   \author{Ioan Kosztin}
   \address{Beckman Institute and Department of Physics, University of
     Illinois, Urbana, IL 61801}

  \date{\today} 

\maketitle

\begin{abstract}
  In this paper we demonstrate how, using a natural generalization of BCS
  theory, superconducting phase coherence manifests itself in phase
  insensitive measurements --- when there is a smooth evolution of the
  excitation gap $\Delta$ from above to below $T_c$. In this context, we
  address the underdoped cuprates. Our premise is that just as Fermi
  liquid theory is failing above $T_c$, BCS theory is failing below.
  The order parameter $\Delta_{sc}$ is different from the excitation gap
  $\Delta$.  Equivalently there is a (pseudo)gap in the excitation
  spectrum above $T_c$ which is also present in the underlying normal
  state of the superconducting phase.  A central emphasis of our paper
  is that the latter gap is most directly inferred from specific heat
  and vortex core experiments.  At the same time there are indications
  that fermionic quasiparticles exist below $T_c$ so that many features
  of BCS theory are clearly present. A natural reconciliation of these
  observations is to modify BCS theory slightly without abandoning it
  altogether.  Here we review such a modification based on a BCS-like
  ground state wavefunction.  A central parameter of our extended BCS
  theory is $\Delta^2 - \Delta_{sc}^2$ which is a measure of the number
  of bosonic pair excitations which have a non-zero net momentum.  These
  bosons are present in addition to the usual fermionic quasiparticles.
  Applying this theory we find that the Bose condensation of Cooper
  pairs, which is reflected in $\Delta_{sc}$, leads to sharp peaks in
  the spectral function once $T \le T_c$. These are manifested in ARPES
  spectra as well as in specific heat jumps, which become more like the
  behavior in a $\lambda$ transition as the pseudogap develops.  We end
  with a discussion of tunneling experiments and condensation energy
  issues.  Comparison between theoretical and experimental plots of
  $C_v$, and of tunneling and vortex core spectroscopy measurements is
  good.
\end{abstract}

\pacs{PACS numbers: 
74.20.-z, %Theories and models of superconducting state
74.20.Fg, % BCS theory and its development<br>
74.25.Bt, %Thermodynamic properties<br>
74.25.Fy %Transport properties (electric and thermal conductivity, 
         %thermoelectric effects, etc.)<br>
}
}

%------------------------------------------------------

\section{Introduction}

In the underdoped regime of high temperature superconductors it is now
clear that Fermi liquid theory is failing and the ``smoking gun" for
this failure is a (pseudo)gap in fermionic excitation spectrum above
$T_c$.  Many would argue\cite{Lee,Anderson} that this failure is
evidence for spin-charge separation. However, the fact that the
excitation gap evolves smoothly into its counterpart in the
superconducting phase may also be interpreted as evidence for
``preformed pairs".  In this way superconducting pairing correlations
are responsible for the breakdown of the Fermi liquid state.  This
picture appears rather natural in view of the notably short coherence
length $\xi$ in these materials, which leads to a breakdown of the
strict mean field theory of BCS.  Within this short $\xi$ scheme, one
considers that pairs form at temperature $T^*$ and Bose condense at
lower temperature $T_c$. These are not true ``preformed" or bound pairs
but rather long lived\cite{Janko,Maly1,Maly2} pair states. Many have
argued for this viewpoint from experimentalists
\cite{Junod,Uemura,Renner,Timusk,Deutscher} to
theorists.\cite{Randeriareview,NSR}

Our contribution\cite{Kosztin1,Kosztin2,Chen1,Chen2,Chen3} to this body
of work has been to show how to microscopically implement this
preformed pair approach at all temperatures $T \le T_c$, by deriving an
extension of BCS theory, based on the ground state of
Leggett.\cite{Leggett} We have also addressed \cite{Janko,Maly1,Maly2}
the behavior above $T_c$.  In this extended BCS approach, the fermionic
excitation gap $\Delta$ (which evolves smoothly from above to below
$T_c$\cite{arpesanl,arpesstanford}) and the mean field order parameter
$\Delta_{sc}$ (which is non-vanishing below $T_c$) are not necessarily
the same at any non-zero temperature; this is a reflection of the
distinction between $T^*$ and $T_c$.  Since $\Delta \ne \Delta_{sc}$, we
say that there are pseudogap effects below $T_c$. \textit{The normal
  state underlying the superconducting phase is not a Fermi liquid}.
The excitations of the system can be viewed as a ``soup" of fermions and
pairs of fermions (bosons). The latter are very long-lived at and below
$T_c$ in the long wave-length limit; their number is associated with the
difference $\Delta^2 - \Delta_{sc}^2$.

This background sets the stage for the important questions which we
address in this paper. What are the signatures of $T_c$, in
thermodynamical quantities such as the specific heat, $C_v$, given the
smooth evolution of the excitation gap?  How do we understand the abrupt
appearance of long lived, fermionic ``quasiparticles" below $T_c$, and
their implications for the electronic spectral function $A ( {\bf k } ,
\omega )$?  If the superconducting state is not Fermi-liquid based, then
how does one extrapolate the ``normal state" below $T_c$ in order to
deduce such thermodynamical properties as the condensation energy?

One of the central observations underlying this paper is the fact that
there are two distinct experiments which provide seemingly similar
information about the extrapolated normal state, i.e., that this $T \le
T_c$ state contains an excitation gap. These are scanning tunneling
microscopy (STM) data in a vortex core\cite{Renner,Davis} as well as
specific heat measurements.\cite{LoramJSC}  Renner and
co-workers\cite{Renner} have argued that their vortex experiments ``show
either the presence of important superconducting fluctuations or
preformed pairs".  Loram\cite{LoramJSC,Loram,Loram98} and co-workers
have analyzed their specific heat data to show that a thermodynamically
consistent picture of $C_v$ reflects a gap in the ($T \le T_c$)
``normal" state spectrum, not directly related to the condensate.  This
would, they argue, include the possibility of preformed pairs that
retained their structure below $T_c$, as in He$^4$, and whose binding
energy does not contribute to the condensation energy.  Alternative
explanations for the vortex core experiments have been advanced by Franz
and Millis,\cite{Franz} and more recently by Franz and
Tesanovic\cite{Tesanovic} and by Lee and Wen.\cite{LeeWen2}

This ``preformed pair" picture, which is essentially
a mean field-based approach,  should be contrasted with the phase
fluctuation picture of Emery and Kivelson\cite{Emery} which has been
implemented by Franz and Millis\cite{Franz} to address photoemission and
tunneling spectroscopies.  In the present case the normal state
excitations represent a ``soup" of fermions and bosons, whereas in the
approach of Ref.~\onlinecite{Franz} the system is thought to consist of
a soup of fluctuating vortices.  It is widely believed that, at least
below $T_c$, fermionic quasiparticles are present so that the phase
fluctuation picture will eventually need to accommodate their
contributions.  Loram\cite{Loram98} has, moreover, argued that phase
fluctuations are not consistent with the behavior of $C_v$, which he
observes.

The results which we obtain in this paper show that upon entering the
superconducting phase, the onset of the coherent condensate,
characterized by $\Delta_{sc}$, leads to a sharpening of the peaks in
the electronic spectral function, which will be directly reflected in
ARPES studies where its effects are quite dramatic, as well as in
tunneling.\cite{note1} ARPES measurements support such a peak
sharpening, and it has been recently claimed,\cite{Shennew} as is
consistent with the theme of this paper, that the observed sharpening at
$T_c$ (rather than at $T^*$) is difficult to understand within strict
BCS theory.

Indeed, BCS theory can and should be generalized, and in its more
general form, this peak sharpening in conjunction with the temperature
dependence of the excitation gap, is also responsible for a specific
heat jump.  The latter, is thus, quite generally, associated with the
onset of off-diagonal long range order.  When this general picture is
applied to the cuprates we find that in the overdoped regime, this jump
(in $C_v$) can be quantified in terms of the temperature dependence of
the excitation gap $\Delta$ (as in the traditional BCS case, see
Eq.~\ref{Cv_Jump} below).  This is in contrast to the underdoped regime,
where the order parameter and excitation gap are distinct and where the
excitation gap is smooth across $T_c$.  Here the jump (associated only
with $\Delta_{sc}$) becomes smaller towards underdoping, where the
pseudogap is more prominent. Moreover, the shape of the $C_v$
\textit{versus} $T$ curve is more like the $\lambda$ transition of
Bose-Einstein condensation (BEC).  All of these features seem to be
consistent with experiment.\cite{Loram98,Junod}

An important second theme of this paper is an analysis of the
extrapolated normal state below $T_c$.  We argue here that the
superconductivity is non-Fermi liquid based and that this has important
implications for condensation energy estimates. Moreover, the non-Fermi
liquid characteristics of the extrapolated normal state (which underlies
the superconducting phase) should help provide constraints on the long
standing controversy\cite{Anderson} of Fermi liquid break-down in the
normal state. Indeed, spin-charge separation
scenarios\cite{LeeWen2,Tesanovic} might be distinguishable from
alternatives such as the present one, or that of Ref.\onlinecite{Franz}
by studies \textit{below} $T_c$.  This provides a major impetus for the
present work.

\section{Theoretical Framework}

\subsection{Overview}

Our work begins with the ground state wavefunction of BCS
as generalized by Leggett\cite{Leggett}
\begin{equation}
\Psi_0 = \Pi_{\mathbf{k}} (
u_{\mathbf{k}} + v_{\mathbf{k}} c^\dagger_{\mathbf{k}\uparrow}
c^\dagger_{\mathbf{-k}\downarrow} )|0\rangle 
  \label{eq:1}
\end{equation}
which describes the continuous evolution between a BCS system, having
weak coupling $g$ and large $\xi$, towards a BEC system with large $g$
and small $\xi$.  Here $ u_{\mathbf{k}}, v_{\mathbf{k}}$, which are
defined as in BCS theory, are self consistently determined in
conjunction with the number constraint.  \textit{The central
  approximation of this paper is the choice of this ground state
  wavefunction}.  The essence of our previous
contributions\cite{Kosztin1,Chen2,Chen3} has been a characterization of
the excitations of $\Psi_0$ and their experimental signatures (for all
$T \le T_c$).  New thermodynamical effects stemming from bosonic degrees
of freedom must necessarily enter, as one crosses out of the BCS regime,
towards Bose-Einstein condensation.

As in BCS theory, we presume that there exists some attractive
interaction between fermions of unspecified origin which is written as
$V_{\mathbf{k,k'}} = g\varphi_{\mathbf{k}} \varphi_{\mathbf{k'}}$, where
$g<0$; here, $\varphi_{\mathbf{k}}=1$ and $(\cos k_x -\cos k_y)$ for
$s$- and $d$-wave pairing, respectively.  The fermions are assumed to
have dispersion, $\epsilon_{\mathbf{k}}=2t_\parallel (2-\cos k_x -\cos
k_y) + 2t_\perp (1-\cos k_\perp) -\mu$, measured with respect to the
fermionic chemical potential $\mu$. Here $t_\parallel$ and $t_\perp$ are
the in-plane and out-of-plane hopping integrals, respectively. In a
quasi-two dimensional (2D) system, $t_\perp \ll t_\parallel$. For
brevity, we use a four momentum notation $K\equiv(\mb{k}, i\omega)$,
$\sum_K \equiv T\sum_{\mb{k}, \omega}$, etc., and suppress $\varphi_{\bf
  k}$ until the final equations.

We now make a number of important observations
about BCS theory.  BCS theory involves a special
form for the pair susceptibility $ \chi (Q) = \sum_K G (K) G_0 (Q-K) $,
where the Green's function $G$ satisfies $G^{-1} = G_0 ^{-1} + \Sigma $,
with $ \Sigma(K) = -\Delta_{sc}^2 G_0(-K)$.  In this notation, the gap
equation is
\begin{equation} 
  1 + g \chi (0) = 0,\qquad T \le T_c.
  \label{eq:gap0} 
\end{equation}  
As was first observed by Kadanoff and Martin,\cite{Kadanoff} this BCS
gap equation can be rederived by truncating the equations of motion so
that only the one ($G$) and two particle (${\cal T}$) propagators
appeared.  Here $G$ depends on $\Sigma$ which in turn depends on $ {\cal
  T} $. In general ${\cal T}$ has two additive
contributions,\cite{Kosztin1} from the condensate (sc) and the
non-condensed (pg) pairs. Similarly the associated self
energy\cite{Kadanoff}

\begin{equation}
  \Sigma (K) = \sum_Q {\cal{T}} (Q) G_0(Q-K)
\end{equation}
can be decomposed into $\Sigma_{pg}(K) + \Sigma_{sc}(K)$.  The two
contributions in $\Sigma$ come respectively from the condensate,
${\cal{T}}_{sc}(Q) = -\Delta_{sc}^2 \delta (Q)/T$, and from the $Q \ne
0$ pairs, with ${\cal{T}}_{pg}(Q) = g/(1 + g \chi (Q))$.

More generally, at larger $g$, the above equations hold but we now
include feedback into Eq.~(\ref{eq:gap0}) from the finite momentum
pairs, via $\Sigma_{pg} (K) = \sum_Q {\cal{T}}_{pg}(Q) G_0 (Q- K)
\approx G_0 (-K) \sum_Q {\cal{T}}_{pg} (Q) \equiv -\Delta_{pg}^2
G_0(-K)$, which defines a pseudogap parameter, $\Delta_{pg}$.  This last
approximation is valid only because (through Eq.~(\ref{eq:gap0})),
${\cal{T}}_{pg}$ diverges as $Q \rightarrow 0$.  In this way,
$\Sigma_{pg} (K)$ has a BCS-like form, as does the total self energy
$\Sigma (K) = - \Delta^2 G_0(-K)$, where
\begin{equation}
  \Delta ^2 = \Delta_{sc}^2 + \Delta_{pg}^2.
  \label{eq:2}
\end{equation}

For the physically relevant regime of moderate $g$, we have found, after
detailed numerical calculations,\cite{Chen1,Chen2} that ${\cal T}_{pg}$
may be approximated as
\begin{equation} 
  {\cal T}_{pg}^{-1}({\mathbf{q}}, \Omega)= a_0 (\Omega -
  \Omega_{\mathbf{q}} + \mu_{pair}+ i\Gamma_{\mathbf{q}}) \:.
  \label{eq:4}
\end{equation}
where the pair dispersion $\Omega_{\mathbf{q}} = q^2/2 M_{pair}$ and
the effective pair chemical potential $\mu_{pair} = 0$ for $T \le
T_c$. The effective pair mass $M_{pair}$ and the coefficient $a_0$ are
determined via a Taylor expansion\cite{Chen3} of ${\cal T}_{pg}^{-1}$.
Moreover, $\Gamma_{\bf q} \rightarrow 0$, as ${\bf q} \rightarrow 0$.
As a consequence we have
\begin{equation}
  \label{eq:gap3}
   \Delta_{pg}^2 = -\sum_Q {\cal{T}}_{pg}(Q) =
\frac{1}{a_0} \sum_{ {\bf q} \ne 0}  b (\Omega_{\bf q}).
\end{equation}
We now rewrite Eq.~(\ref{eq:gap0}), along with the fermion number
constraint, as
\begin{eqnarray}
  \label{eq:gap1}
  1+g\sum_{\mathbf{k}} \frac{1- 2f(E_{\mathbf{k}})}{2E_{\mathbf{k}}}\,
  \varphi^2_{\mathbf{k}} &=& 0\;, \\ 
 \label{eq:gap2}
 \sum_{\mathbf{k}}\left[1- \frac{\epsilon_{\mathbf{k}}}{E_{\mathbf{k}}} +
   \frac{2\epsilon_{\mathbf{k}}}{E_{\mathbf{k}}}\,f(E_{\mathbf{k}})\right]
 &=& n \end{eqnarray} 
Here $f(x)$ and $b(x)$ are the Fermi and Bose functions and
$E_{\mathbf{k}} = \sqrt{ \epsilon_{\mathbf{k}} ^2 + \Delta^2
  \varphi_{\mathbf{k}}^2}$ is the quasiparticle dispersion.  

Equations (\ref{eq:gap3})-(\ref{eq:gap2}) represent the central
equations of our theory below $T_c$. They are consistent with BCS theory
at small $g$, and with the ground state $\Psi_0$ at all $g$; in both
cases the right hand side of Eq.~(\ref{eq:gap3}) is zero. The simplest
physical interpretation of the present decoupling scheme is that it goes
beyond the standard BCS mean field treatment of the single particles
(which also acquire a self energy from the finite ${\bf q}$ pairs), but
it \textit{treats the pairs at a self consistent mean field level}.

\section{Spectral Functions, Densities of States and Specific Heat}
\label{Sec_Spectral}

Experimentally, it has been established from specific heat measurements
in the cuprates that there is a step discontinuity or a maximum at
$T_c$, depending on the doping level. \cite{Loram,Loram98} It is clear
that one cannot explain these experiments using the standard picture of
BCS theory, in which the specific heat jump at $T_c$ results from the
opening of the excitation gap. We now address these experiments.  In
Section II, we used an approximate form for the pseudogap self-energy
$\Sigma_{pg}$ (see derivation of Eq.~(\ref{eq:2})), in order to simplify
the calculations.  Under this approximation, $\Sigma_{pg}$ has a
BCS-like character, so that the spectral function is given by two
$\delta$-functions at $\pm \Ek$.  These approximations were justified in
the context of the applications considered, thus
far.\cite{Kosztin1,Kosztin2,Chen1,Chen2,Chen3}

However, in order to study quantities which rely on details of the
density of states, we will, in the remainder of this paper, relax this
simplifying approximation and allow for lifetime effects in
$\Sigma_{pg}$. This more realistic form for $\Sigma_{pg}$ incorporates a
finite broadening $\gamma$ due to the incoherent nature of the finite
center-of-mass momentum pair excitations.  In this way we distinguish
this contribution from that of the condensate. To make numerical
calculations tractable,\cite{Maly1,Maly2} we do not solve for the
broadening $\gamma$ and excitation gap $\Delta$ self-consistently. This
would involve an iterative solution of the complex set of three coupled
equations for $G$ and ${\cal{T}}$; rather we take $\Delta$ below $T_c$
from Eqs.~(\ref{eq:gap3})- (\ref{eq:gap2}), and use our
estimates\cite{Maly1,Maly2} of $T^*$ to determine where pseudogap
effects are essentially negligible.  More importantly, we treat the
broadening as a phenomenological parameter, with one adjustable
coefficient, chosen to optimize fits to tunneling experiments.  Our
results, throughout this paper, are not particularly sensitive to the
detailed form of $\gamma$ (which will depend on doping concentration $x$
and $T$),\cite{gamma_note} but it is essential that $\gamma$ be non-zero
and appreciable when compared with $T$.  In this same spirit, we take
$T_c$, and the chemical potential $\mu$ from our leading order
calculations (with $\gamma$=0).

We turn now to the spectral function $A(\mb{k}, \omega)$.  It follows
from our microscopic scheme that slightly above\cite{Maly1,Maly2}
$T_c$, and for all $T \le T_c$,\cite{Kosztin1} the self energy
associated with the ${\bf q } \ne 0 $ pairs, and that from the
condensate or ${\bf q } = 0 $ pairs is given respectively by
\begin{equation}
\Sigma_{pg}(\mb{k},\omega) =
\frac{\Delta_{\mb{k},pg}^2}{\omega+\ek+i\gamma} -i\Sigma_0 (\mb{k}, \omega) \:,
\label{SigmaPG_Model_Eq} 
\end{equation}
and
\begin{equation}
\Sigma_{sc}(\mb{k},\omega)=
\frac{\Delta_{\mb{k},sc}^2}{\omega +\ek} \:,
\label{SigmaSC}
\end{equation}
where $\Delta_{\mb{k},pg}=\Delta_{pg}\phik$ and
$\Delta_{\mb{k},sc}=\Delta_{sc}\phik$. Here we have added to
$\Sigma_{pg}$ an additional piece $\Sigma_0$ which is not accounted for
by our (particle-particle) ladder diagrams.  This leads to an
``incoherent" background contribution which we will address later in the
context of the cuprates. For the rest of the discussion in the next two
sections we set $\Sigma_0 =0$ and, thereby, focus only on the
superconducting and pseudogap terms.

\begin{figure*}[t]
\centerline{\includegraphics[width=5.5in, clip]{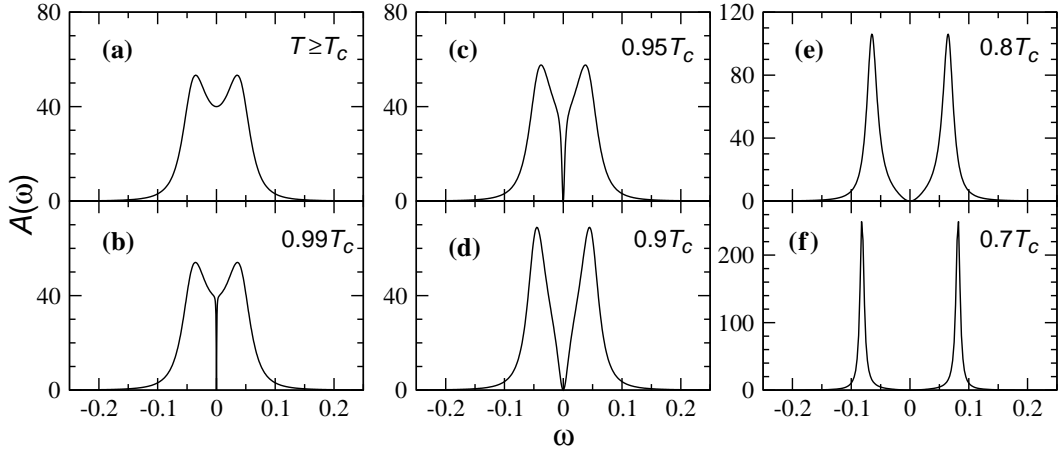}}
\smallskip
\caption{Behavior of the coherent contribution to the spectral function
at various $T$, for wavevectors away from the nodes.}
\label{SpectralFunction}
\end{figure*}

The spectral function can readily be computed from $\Sigma = \Sigma_{sc} +
\Sigma_{pg}$, as
 \begin{equation} 
A(\mb{k},\omega)=-2\,\mbox{Im}\, G(\mb{k},\omega+i0) \:,
  \label{Spectral_Eq}
\end{equation}
which satisfies the sum rule ${ \int_{-\infty}^\infty
  \frac{\mbox{d}\omega}{2\pi} A(\mb{k},\omega) = 1 }$.  We, thus, obtain
a relatively simple expression for $A(\mb{k},\omega)$ which applies
below and above $T_c$, respectively:
\begin{mathletters}
\begin{eqnarray}
A(\mb{k},\omega) &=&\frac{2\Delta_{\mb{k},pg}^2 \gamma (\omega+\ek)^2}
{(\omega+\ek)^2 (\omega^2-\Ek^2)^2 + \gamma^2
  (\omega^2-\ek^2-\Delta_{\mb{k},sc}^2)^2} \:,\nonumber\\ 
\label{SpectralF_Eq_1}
\\
A(\mb{k},\omega) &=& \frac{2\Delta_{\mb{k}}^2\gamma} {(\omega^2-\Ek^2)^2
+\gamma^2(\omega-\ek)^2} \:,
\label{SpectralF_Eq}
\end{eqnarray}
\end{mathletters}
From Eq.~(\ref{SpectralF_Eq_1}), we see that the spectral function
contains a zero at $\omega=-\ek$ below $T_c$, whereas it has no zero
above $T_c$. This difference is responsible for the different
thermodynamical behavior across $T_c$.

In Fig.~\ref{SpectralFunction}, we plot the spectral function for
$\ek=0$ (on the Fermi surface) at different temperatures from slightly
above $T_c$ [Fig.~\ref{SpectralFunction}(a)] to temperatures within the
superconducting phase [Fig.~\ref{SpectralFunction}(f)]. This figure is
typical of situations in which there is a well established pseudogap.
The figure can be viewed as representative of both $s$- and $d$-wave
order parameter symmetries.  Hence the value of the wave-vector
$\hat{k}$ is not particularly relevant, provided it is away from the
nodal points in the $d$-wave case.  For illustrative purposes, we take
$\hat{k}$ at the anti-nodes, with $\gamma (T) =\Delta_{pg} (T_c)$, and
$\Delta_{pg}(T_c) = 0.05$ (in units of $4t_\parallel$).  In this way we
ignore any $T$ dependence in $\gamma$ and, thus, single out the long
range order effects associated with $\Delta_{sc}$.

These figures give the first clear indications of the onset of
``quasiparticle" coherence. Moreover, panel (a) helps to emphasize an
important component of our physical picture: the superconductor is not
in a Fermi liquid state just above $T_c$, as can be seen by the
non-Fermi liquid form for the spectral function.  Just below $T_c$, the
dramatic dip at $ 0.99 T_c$ is a consequence of Bose condensation of
${\bf q} = 0$ pairs.  Here, a very small condensate contribution
nevertheless leads to the depletion of the spectral weight at the Fermi
level, as shown in Fig.~\ref{SpectralFunction}(b). As the temperature
continues to decrease, and the superconducting gap increases, the two
peaks in the spectral function become increasingly well separated, as
plotted in Figs.~\ref{SpectralFunction}(c)-(f).  Even at the relatively
high temperatures corresponding to $T/T_c \sim 0.7$, the spectral peaks
are quite sharp --- only slightly broadened relative to their BCS
counterparts (where the spectral function is composed of two $\delta$
functions).

It should be stressed that lifetime effects via $\gamma$ do not lead to
significant peak broadening.  This follows from the fact that the
imaginary part of the pseudogap self-energy at the peak location $\Ek$
is given by (for $\epsilon_{\bf k} = 0$)
\begin{equation}
\gamma^\prime = \gamma \frac{\Delta_{\mb{k},pg}^2}{(\Ek+|\ek|)^2+\gamma^2} =
\gamma \frac{\Delta_{\mb{k},pg}^2}{\Delta_\mb{k}^2+\gamma^2} \:.
\label{Gamma'_Eq}
\end{equation} 
Since Eq.~(\ref{eq:gap3}) indicates that $\Delta_{pg}$ vanishes as $T
\rightarrow 0$, the effective peak width, determined by $\gamma^\prime$,
decreases with decreasing $T$.  It can be seen that below $T_c$, the
spectral function in Eq.~(\ref{SpectralF_Eq}), is very different from
that obtained using a simple broadened BCS form; there is no true gap
for the latter, in contrast to the present case.

\begin{figure*}
\centerline{\includegraphics[width=6in, clip]{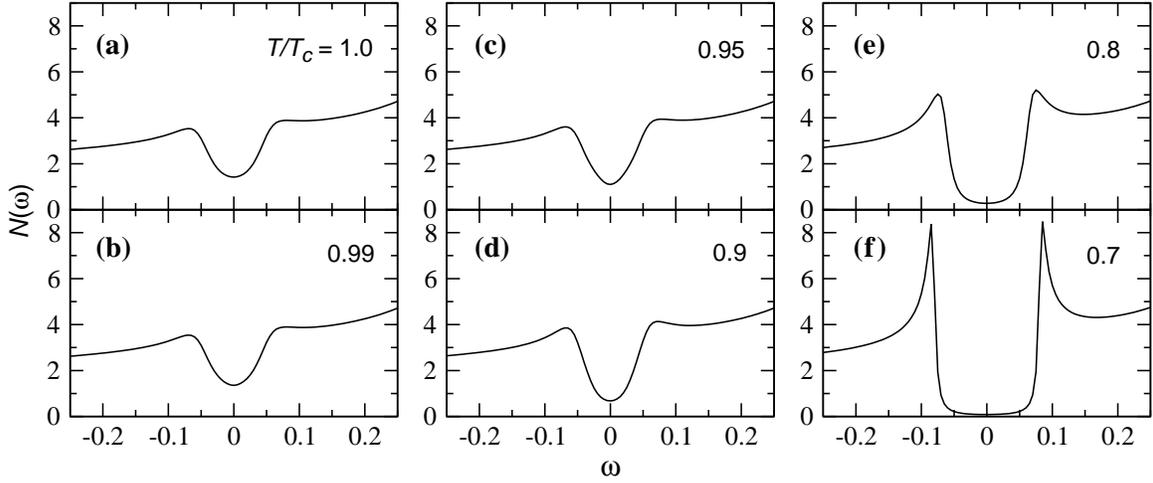}}
\medskip
\caption{Effects of superconducting long range order on
  the behavior of the density of states as a function of temperature in
  a pseudogapped $s$-wave superconductor with $n=0.5$. Here we take the
  parameters to be the same as those used
  in Fig.~\ref{SpectralFunction}.
  At $T/T_c\sim 0.7$, as shown in (f), the density of states is
  close to that of strict BCS theory.}
\label{DOS}
\end{figure*}

These spectral functions can be used to derive the density of states (per
spin) as
\begin{equation}
N(\omega)=\sumk A(\mb{k},\omega) \:.
\end{equation}
Moreover, it is expected that peak sharpening effects discussed above
for the spectral function are also reflected in the density of states.
For simplicity, we first consider the case of $s$-wave pairing.  In
Fig.~\ref{DOS}(a)-(f), the density of states is plotted for a quasi-2D
$s$-wave superconductor, where the various energy gaps are taken to be
the same as in Fig.~\ref{SpectralFunction}.  Because of contributions
from states with $\ek\neq 0$, the narrow dips in
Fig.~\ref{SpectralFunction}(b)-(c) do not show up here. However, as is
evident, the density of states within the gap region decreases quickly,
as the superconducting condensate develops.

The rapid decrease of the density of states with decreasing $T$, in the
vicinity of $T_c$, will be reflected in the behavior of the specific
heat, $C_v$ and, thereby, lead to the thermodynamical signature of the
phase transition.  $C_v$ may be obtained from $C_v=dE/dT$, where the
energy $E$ is calculated  via:\cite{Fetter}
\begin{eqnarray}
  E &=& 2T\sum_{\mb{k},n} \frac{1}{2}(i\omega_n + \ek^0 +
  \mu)G(\mb{k},i\omega_n) \nonumber\\ &=& \sumk \int_{-\infty}^\infty
  \frac{\mbox{d}\omega}{2\pi} (\omega+\ek+2\mu)
  A(\mb{k},\omega)f(\omega) \:,
\label{EnergyInt_Eq}
\end{eqnarray}
where $\ek^0=\ek+\mu$ is the dispersion measured from the bottom of the
band.  It follows that
\begin{equation}
E = \int_{-\infty}^\infty \frac{\mbox{d}\omega}{2\pi}
[(\omega+\mu) N(\omega)+K(\omega)]f(\omega) \:,
\end{equation}
where we have defined $K(\omega)\equiv \sumk \ek^0 A(\mb{k},\omega)$,
which can be regarded as the contribution associated with the kinetic
energy of the system.  In this way, we obtain
\begin{eqnarray} 
C_v &=& \int_{-\infty}^\infty \frac{\mbox{d}\omega}{2\pi} \left\{
  \frac{\partial \mu}{\partial T} N(\omega) f(\omega) \right.\nonumber\\
 &&{}  -\Big[(\omega+\mu)N(\omega)+K(\omega)\Big]\frac{\omega} {T}
  f^\prime(\omega) \nonumber\\ 
&&{} + \left.
  \left[(\omega+\mu)\frac{\partial N(\omega)}{\partial T} + \frac{\partial
      K(\omega)}{\partial T}\right]f(\omega) \right\} \:.  
\label{Cv_Eq}
\end{eqnarray}
The first two terms on the right hand side lead to a ``normal metal-like"
contribution to $C_v/T$ which is proportional to
$N(\omega)$ at low $T$.  However, the third term arises because
$N(\omega)$ depends on $T$.  \textit{In this case, $C_v/T$ no longer
  reflects the density of states.} It is this term that will give rise
to the specific heat discontinuity at $T_c$.

\begin{figure}[t]
  \centerline{\includegraphics[width=3.0in,clip]{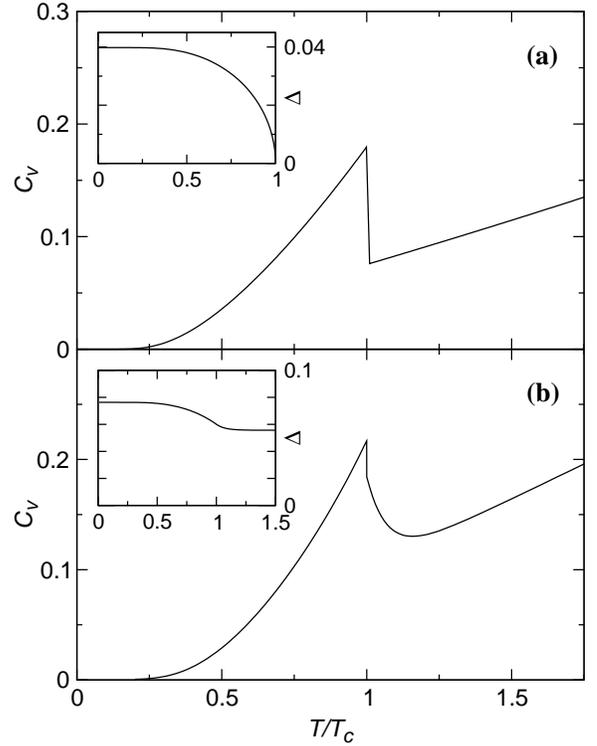}}
\medskip
\caption{Comparison of the temperature dependence of the specific heat
  in the (a) weak coupling BCS case and (b) moderate coupling pseudogap
  case. Shown here are quasi-2D $s$-wave results, at $n=0.5$,
  $-g/4t_\parallel = 0.5$ and 0.6, respectively. The $T$ dependence of
  the gap is shown as insets.  }
\label{Cv_S}
\end{figure}

In Fig.~\ref{Cv_S} we plot the temperature dependence of $C_v$ in both
(a) the weak coupling BCS case and (b) the moderate coupling pseudogap
case with $s$-wave pairing. We choose, for definiteness, the broadening
$\gamma(T)=T$ for the second of these calculations.  We also indicate in
the insets, the respective temperature dependent excitation gaps, which
have been assumed\cite{Kosztin1,Chen2} in producing the figure.

In both cases shown in Fig.~\ref{Cv_S}, the specific heat jump arises
from a discontinuity in $\mbox{d}N(\omega)/\mbox{d}T$,\cite{newfootnote}
associated with the onset of superconducting order.  However, for the
BCS case, this derivative can be associated with a discontinuity in the
derivative of the excitation gap, via
\begin{equation}
 \Delta C_v^{BCS} = -N(0) \frac{\mbox{d}\Delta^2} {\mbox{d}T\;}\:.
\label{Cv_Jump}
\end{equation}
By contrast, in the pseudogap case, the gap $\Delta$ and its derivative
$\mbox{d}\Delta/\mbox{d}T$ are presumed to be continuous across $T_c$ as
shown in the inset to Fig.~\ref{Cv_S}(b), and in Fig.~\ref{Cv-Gaps}
below, so that Eq.~(\ref{Cv_Jump}) does not hold.  Moreover, in this
case, above but near $T_c$, the temperature dependence in the density of
states is still important due to the presence of an excitation gap above
$T_c$.  The latter leads to a decrease in $\mbox{d}N(\omega)/\mbox{d}T$
which is then reflected \cite{note2} in a decrease in $C_v$, slightly
above $T_c$.  At higher $T$, well away from $T_c$, where
$\mbox{d}N(\omega)/\mbox{d}T$ tends gradually to zero, $C_v$ is then
controlled, as in a more typical ``normal metal", by $N(\omega)$.  We
see, then, that the approach to the ``normal metal" value is sharp for
BCS, but because of the non-zero pseudogap, it is more gradual for case
(b).  An important consequence of these effects, is that the shape of
the anomaly in $C_v$ (shown in Fig.~\ref{Cv_S}(b)) is more
characteristic of a $\lambda$-like transition, although there is a
precise step function discontinuity just at $T_c$.

\section{Application to the cuprates}
\label{Sec_Appl_Cuprates}

The results obtained in Sec.~\ref{Sec_Spectral} are generally valid for
both $s$- and $d$-wave cases, and can be readily applied to the $d$-wave
cuprates.  In this section we test the physical picture and the results
obtained above, by studying the tunneling spectra and the specific heat
behavior in the cuprates, as a function of doping and of temperature.
As in earlier work,\cite{Chen2,Chen3} we introduce a hole concentration
dependence of the electronic energy scales by imposing the Mott
constraint that the in-plane hopping integral $t_{\parallel}(x) = t_0
x$, so that the plasma frequency vanishes as $x \rightarrow 0$.  As a
result, the effective coupling strength $-g/t_{\parallel}(x) $ increases
as the Mott insulator phase is approached.  Here we assume (for
simplicity) $g(x) = g$ and fit the one free parameter $g/t_0$ to the
phase diagram.\cite{Chen2}

\begin{figure}
  \centerline{\includegraphics[width=3.in,clip]{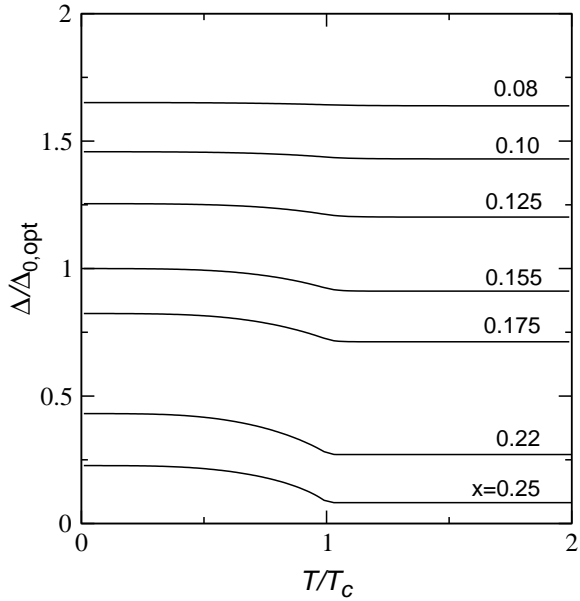}}
\medskip
\caption{Temperature dependence of the excitation gaps for various
  doping concentrations used for calculations in Fig.~\ref{Cv-x}. Here
  $\Delta_{0, opt}$ is the zero $T$ gap at optimal doping $x\approx 0.15$.}
\label{Cv-Gaps}
\end{figure}

In order to compare with tunneling spectra, we introduce a slightly more
realistic band structure which includes a next-nearest neighbor hopping
term, $-2t^\prime (1-\cos k_x \cos k_y)$, in the band dispersion, $\ek$,
with $t^\prime/t_\parallel \approx 0.4$. This parameter choice gives
rise to the hole-like Fermi surface shape seen in ARPES measurements for
under- and optimally doped cuprates,\cite{Ding,note3} and places the van
Hove singularity in a more correct position within the band.
 
Finally, we turn to the phenomenological parameter $\gamma$ as well as
to $\Delta$.  We presume that $\gamma$ changes from above to below $T_c$
and in this way $\Delta$ (which is directly coupled to $\gamma$ via the
set of coupled equations for $G$ and ${\cal{T}}$ ) will have some,
albeit small, structure in its temperature dependence at $T_c$, as seems
to be the case experimentally.  Our choice for the excitation gaps is
shown in Fig.~\ref{Cv-Gaps}, and appears compatible with Figs.~8 and 9
in Ref.~\onlinecite{LoramJSC}.  As is consistent with scattering rate
measurements in the literature,\cite{Hardy2,Valla} we take $\gamma
\propto T^3$ below $T_c$ and linear in $T$ above $T_c$.\cite{Power} For
the doping dependence, we assume that $\gamma$ varies inversely with
$\Delta$. This reflects the fact that when the gap is large, the
available quasiparticle scattering decreases.  With these reasonable
assumptions, along with the continuity of $\gamma$ at $T_c$, we obtain a
simple form:
\begin{equation}
\gamma = \left\{ \begin{array}{l@{\hspace{1cm}}l}
a T^3/T_c\Delta \:, & (T< T_c),\\
a TT_c/\Delta \:, & (T> T_c). \end{array} \right.
\label{Gamma_Model_Eq}
\end{equation}
Here, the coefficient $a \le 1 $. This corresponds to our single
adjustable parameter.

\subsection{Tunneling spectra}
\label{Subsec_Tunneling}

Tunneling experiments were among the first to provide information about
the excitation gap --- which measurements seem to be
consistent with ARPES data.\cite{Timusk}  For a given density of states
$N(\omega)$, the quasiparticle tunneling current across a
superconducting-insulator-normal (SIN) junction can be readily
calculated, \cite{Mahan}
\begin{equation}
I_{SIN} = 2eN_0 T^2_0 \int_{-\infty}^\infty \frac{\mbox{d}\omega}{2\pi}
N(\omega)  \left[ f(\omega-eV)-f(\omega)\right] \:,
\end{equation}
where we have assumed a constant density of states, $N_0$, for the
normal metal, and presumed that the tunneling matrix element $T_0$ is
isotropic. In reality, there may be some directional tunneling which
will tend to accentuate the gap features, but we do not complicate our
discussion here with these effects.  At low $T$, one obtains
\begin{equation}
\left(\frac{\mbox{d}I}{\mbox{d}V}\right)_{SIN} \approx
\frac{e^2N_0T_0^2}{\pi} N(eV) \:
\end{equation}
so that the tunneling spectra and the density of states are equivalent,
up to a multiplicative constant coefficient.  At $T$ comparable to
$T_c$, however, the tunneling spectra reflect a more thermally broadened
density of states.

\begin{figure*}
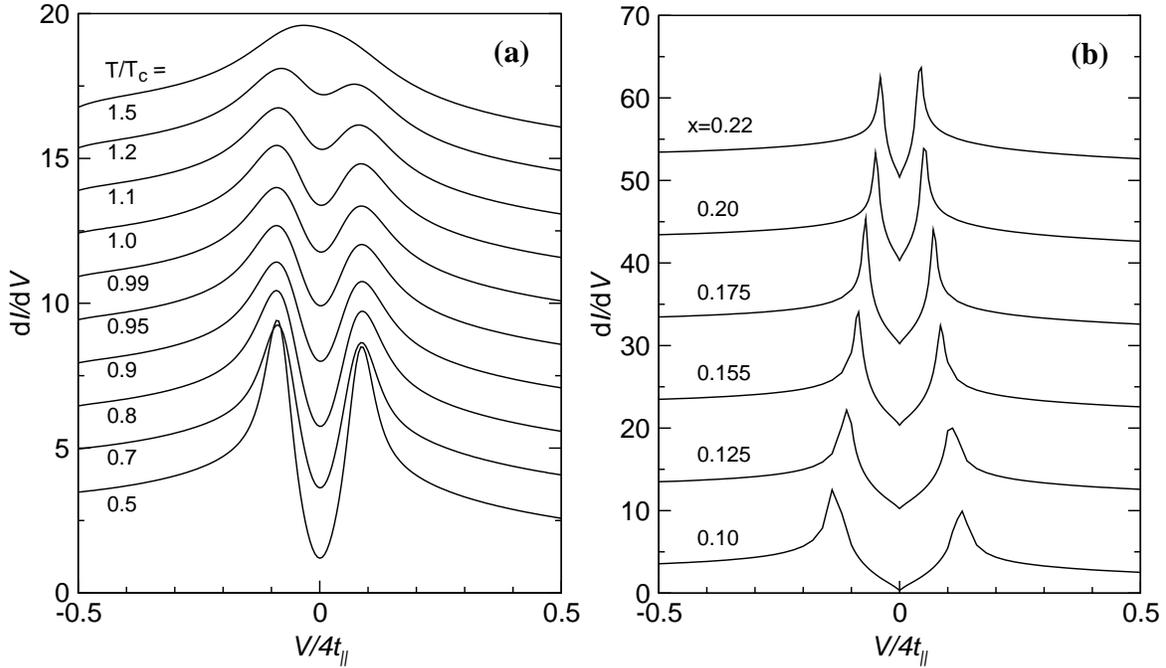

\centerline{\includegraphics[width=3in,clip]{Fig5a}
\includegraphics[width=3in,clip]{Fig5b}}
\medskip
\caption{(a) Temperature and (b) doping dependence of tunneling
  spectra across an SIN junction. Shown in (a) are the
  $\mbox{d}I/\mbox{d}V$ characteristics calculated for optimal doping
  at various temperatures from above to below $T_c$.  Shown in
  (b) are tunneling spectra at low $T$ (at $0.2T_c$) for various $x$.
  The units for $\mbox{d}I/\mbox{d}V$ are $e^2N_0T_0^2/4t_\parallel$,
  where $t_{\parallel}$ is evaluated at optimal doping.
  For clarity, the curves in (a) and (b) are vertically offset by 1.5
  and 10, respectively.}
\label{SIN}
\end{figure*}

In Fig.~\ref{SIN}(a), we plot the SIN tunneling spectra, calculated for
optimal doping ($x \approx 0.15$) at temperatures varying from above to
below $T_c$. The van Hove singularity introduces a broad maximum in the
spectra at high temperatures, as seen for the top curve in
Fig.~\ref{SIN}(a).  We see here that (even for this optimal sample), as
observed experimentally,\cite{Timusk} the density of states contains
(pseudo)gap like features which lead to two peaks. This is visible for
temperatures well above $T_c$.  A similar plot is presented in
Fig.~\ref{SIN}(b), which shows at fixed low $T = 0.2 T_c$ how the
spectrum evolves as a function of $x$. Both these plots appear in
reasonable agreement with what is observed experimentally by Renner and
coworkers \cite{Renner_Tunneling} and by Miyakawa \textit{et al}
\cite{JohnZ} for Bi$_2$Sr$_2$CaCu$_2$O$_{8+\delta}$ (Bi2212).

\begin{figure*}
  \centerline{\includegraphics[width=6in,clip]{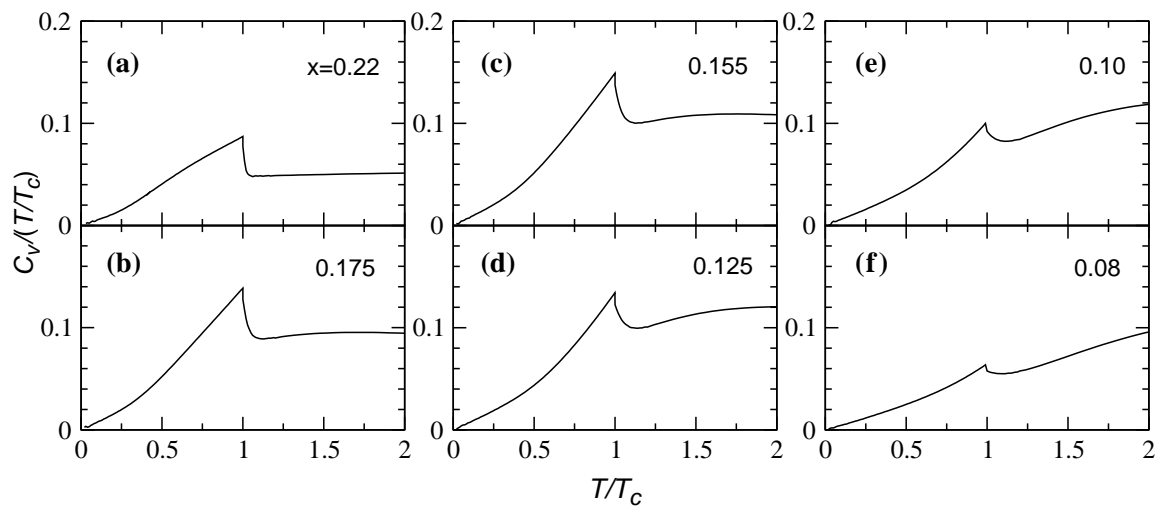}}
\caption{Temperature dependence of the specific heat for various
  doping concentrations, calculated with $a=1/4$ in
  Eq.~\ref{Gamma_Model_Eq}.}
\label{Cv-x}
\end{figure*}

\subsection{Specific heat}
\label{Subsec_Cv}

There is a substantial amount of experimental data on the specific heat
in the cuprates,\cite{Loram98,Junod} although systematic studies come
primarily from one experimental group. \cite{Loram98}
We compare our numerical results with these data by plotting our
calculations for $C_v/T$ in Fig.~\ref{Cv-x}(a)-(f), from over- to
underdoped systems.  As shown in these plots, the behavior of $C_v$ is
BCS-like in the overdoped regime.  As the system passes from optimal
doping towards underdoping, the behavior is more representative of a
$\lambda$-like anomaly, as found in Fig.~\ref{Cv_S}(b) for the $s$-wave
case.  All these trends seem to be qualitatively consistent with
experimental data.\cite{Loram,Loram98}  In the underdoped regime at
high $T$, we find a maximum in $C_v/T$, near a temperature, $T^*$, which
may be associated with the onset of the pseudogap state [See, e.g,
Fig.~\ref{Cv_Extrapolation}(c)].  Finally, in contrast to
Fig.~\ref{Cv_S}, at low $T$, the $d$-wave nodes lead to a larger
quasiparticle specific heat, than for the $s$-wave case.

The experimentally observed $\lambda$-like anomaly of $C_v$ at $T_c$ has
been interpreted previously as evidence for a Bose condensation
description.\cite{Junod} Here, in contrast, we see that within our
generalized mean field theory, this anomaly naturally arises from the
temperature dependence of the fermionic excitation gap which has some
structure at, but persists above, $T_c$, as shown in Fig.~\ref{Cv-Gaps}.
Thus, this is a property of superconductors which have a well
established pseudogap.  That the experimental data, (which, except at
extremely reduced hole concentrations), show a reasonably sharp
($\lambda$-like) structure at $T_c$ --- seems to reinforce the general
theme of this work --- that corrections to BCS theory may be reasonably
accounted for by an improved mean field theory, rather than by, say,
including order parameter fluctuation effects.

\section{Low \textit{T} extrapolation of the pseudogapped normal state}
\label{Sec_Extrapolation}

\subsection{Non-Fermi liquid based superconductivity}

The character of the extrapolated ($T \le T_c$) ``normal state" is at
the core of many topical issues in high $T_c$ superconductivity.
Understanding this state may shed light on the nature of Fermi liquid
break-down above $T_c$.  Moreover, the thermodynamics of this
extrapolated phase provide a basis for estimates of the condensation
energy. This is deduced\cite{Loram98} by integrating the difference
between the entropy of the superconducting state and that of the
extrapolated normal state with respect to $T$.  This extrapolated normal
state also appears as a component of the free energy functional of
conventional Landau-Ginzburg theory.  Indeed, considerable attention has
been paid recently to the condensation energy in the context of
determining the pairing ``mechanism" in the high temperature
superconductors. \cite{Demler,Norman2,Leggettmechanism}  In this regard
what is needed is the difference between the various ``normal" and
superconducting state polarizabilities (e.g., magnetic and electric)
which are thought to be responsible for the pairing. It is important to
stress that \textit{the ``normal state" used in computing the
  polarizabilities should contain an excitation gap which is compatible
  with that found in the experimental data analysis\cite{Loram98} with
  which the microscopically deduced condensation energy is compared}.

We have emphasized that within our physical picture there is an
underlying (pseudo)gap in the normal state below $T_c$. A similar
picture was independently deduced phenomenologically from specific heat
and magnetic susceptibility measurements by Loram and
co-workers\cite{LoramJSC} who arrived at equations like those of Section
II [Eqs.~(\ref{eq:2}) and (\ref{eq:gap1})].  However, they did not
impose a self consistent condition on $\Delta_{pg}$ as in
Eq.~(\ref{eq:gap3}).  Rather the quantity $\Delta_{pg}$ (which they
call $E_g$) is assumed to be $T$-independent.

\begin{figure*}
  \centerline{\includegraphics[width=5.5in,clip]{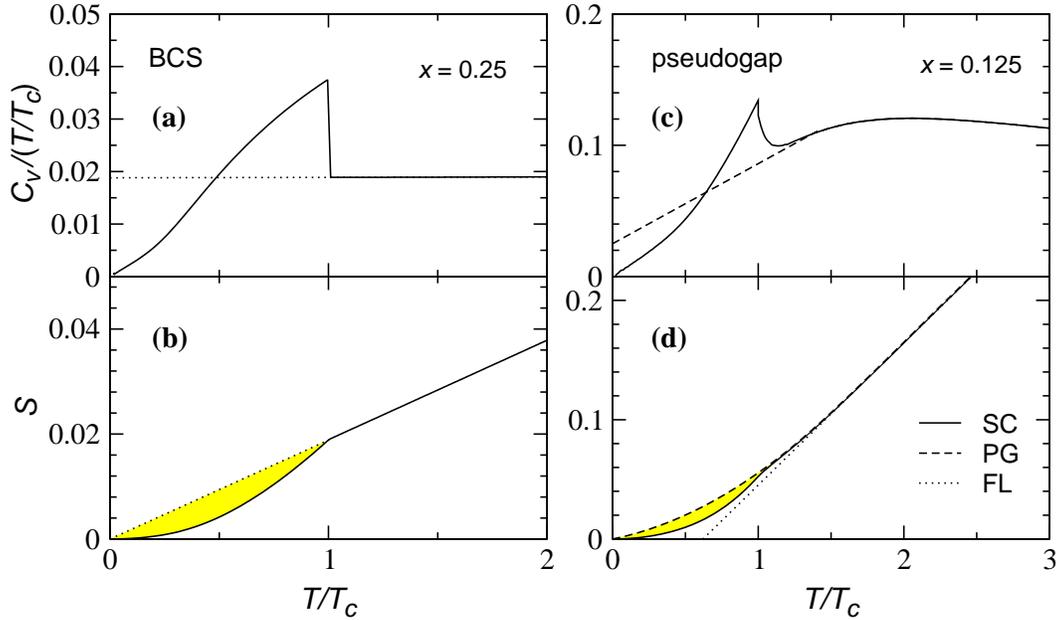}} 
\medskip
\caption{Comparison of the extrapolated normal state
  below $T_c$ in (a)-(b) BCS and (c)-(d) pseudogap superconductors.
  Shown are the extrapolations for $C_v/T$ and the entropy $S$ in the
  upper and lower panels, respectively. Here ``SC'', ``PG'', and ``FL''
  denote superconducting state, extrapolated normal state with a
  pseudogap, and Fermi liquid based extrapolation, respectively. The
  shaded areas in (c) and (d) determine the condensation energy.}
\label{Cv_Extrapolation}
\end{figure*}

To expand on these issues we plot in Fig.~\ref{Cv_Extrapolation} the
calculated $C_v/T$ and entropy $S$ for (a)-(b) the BCS case, as compared
with the counterparts obtained for the pseudogap superconductor in (c)
and (d).  The dotted lines represent the Fermi liquid, i.e., linear
extrapolation (FL). Figures \ref{Cv_Extrapolation}(a) and (b) re-affirm
that this Fermi liquid extrapolation is sensible for the BCS case ---
$C_v/T$ is a constant, and $S$ is a straight line going through the
origin.  Panel (b) is useful in another respect: it shows how the
entropy behaves as phase coherence is established.  In general, the
phase coherent state has a lower entropy than the extrapolated normal
state.

In the underdoped regime, Loram and co-workers\cite{Loram98} have
stressed that entropy measurements lead one to infer that an excitation
gap occurs \textit{above $T_c$}.  We analyze our calculated form of the
entropy in a similar fashion.  In contrast to the BCS case, for a
pseudogap superconductor, the Fermi liquid extrapolation of $S$ is
unphysical, approaching a negative value at low $T$, as shown by the
dotted line in Fig.~\ref{Cv_Extrapolation}(d).  Here the solid and
dotted lines separate around the temperature $T^* $ which we find to be
around $1.5 T_c$.  In order to obtain a thermodynamically consistent
picture, then, the normal state must deviate from the FL line and this
is accomplished by turning on an excitation gap at $T\le T^*$.

The dashed lines in Fig.~\ref{Cv_Extrapolation}(c)-(d) show a more
reasonable extrapolated normal state (labelled PG) which is equivalent
to the solid line for $T \ge T_c$ and distinct for $ T < T_c$. This
extrapolation is taken to be consistent with the conservation of
entropy, $S=\int_0^T C_v/T \mbox{d}T$, i.e., the shaded areas in (c) and
(d). This construction for the $ T \le T_c$ normal state, is similar to
the procedure followed experimentally,\cite{Loram98} and, in effect,
removes phase coherent contributions which enter via $\Delta_{sc}$.
This construction is by no means unique; all that is required is that
the entropy of the extrapolated normal and of the superconducting states
be equal at $T = T_c$. As shown in the figure, we chose, for simplicity,
a straight line extrapolation for $C_v/T$.  Moreover, this choice is
consistent with our expectation that there would be, in the ``normal
state", a finite intercept for $C_v/T$.

Just as a gap is present above $T_c$, the underlying normal phase below
$T_c$ (labelled PG) is to be distinguished from the FL extrapolation; it
also contains an excitation gap.  Indeed, this is consistent with what
has been claimed experimentally:\cite{Loram98} \textit{a
  thermodynamically consistent picture of $C_v$ reflects a gap in the $
  T \le T_c$ normal state spectrum, not directly related to the
  condensate}.  While Figs.~\ref{Cv_Extrapolation}(c)-(d) are similar to
what is in the data in underdoped cuprates, in our analysis the
``normal" state $C_v/T$ and $S$ are linear and quadratic in $T$,
respectively.  Slightly different power laws have been assumed
experimentally.  A rough estimate of the condensation energy can be
obtained from the integrated area between the solid line (for the
superconducting state) and the dashed line (for the extrapolated normal
state) in Fig.~\ref{Cv_Extrapolation}(d).  It should also be noted that
a more meaningful measure of the condensation energy is obtained by
computing the magnetic field dependent Gibbs free energy.  This is more
complicated to implement both theoretically and experimentally.

\subsection{Comparison with vortex core and $C_v$ measurements}

The presence of a pseudogap in the underlying normal state of the
superconducting phase is also consistent with the observations by Renner
and co-workers\cite{Renner} based on STM measurements within a vortex
core. While one might be concerned about magnetic field, $H$, effects in
interpreting these data, it should be noted that $H$ appears to have a
rather weak effect\cite{Pennington} on pseudogap phenomena, as measured
by $T^*$ and $\Delta(H)$.  [In more overdoped samples the field
dependence becomes more apparent\cite{newNMR}]. By contrast $T_c$ is
more sensitive to $H$.

Indeed, this weak dependence on $H$ is often invoked in the literature as
strong evidence against the ``preformed" pair scenario.  In the usual
BCS case, pairs form precisely when phase coherence sets in.  However,
in the case of a pseudogap superconductor where the coupling is
stronger, pairs form above $T_c$ without an underlying phase coherence.
It is clear that a magnetic field lowers $T_c$ by destroying phase
coherence.  However, ``preformed" pairs will survive $H$, leaving the
excitation gap in tact.  Stated alternatively, a magnetic field (just
like magnetic impurities) breaks time reversal symmetry, and therefore
makes it energetically unfavorable to form Cooper pairs which are
comprised of time reversed single particle states. In contrast, finite
momentum pair excitations, which are responsible for the pseudogap in
our approach, are not formed in time reversed states, and as such, they
are not as susceptible to external magnetic fields or to magnetic
impurities.

\begin{figure*}[t]
\centerline{\includegraphics[width=2.5in, clip]{Fig8a}
\hskip 1ex \includegraphics[width=2.5in]{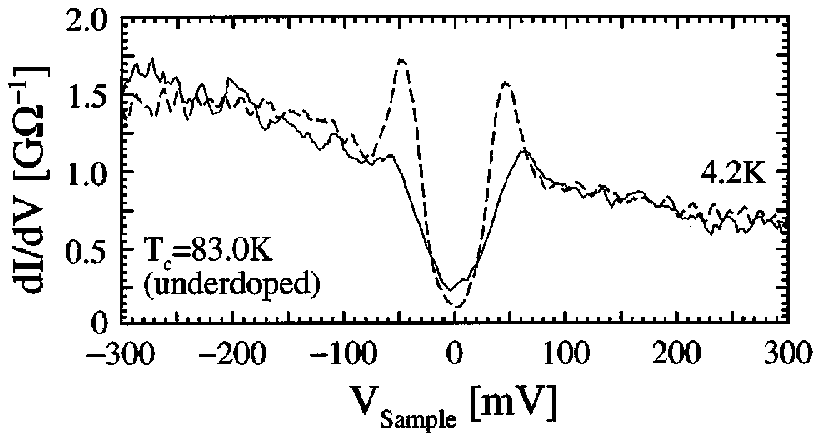}}
\vskip 2ex
\centerline{\includegraphics[width=2.5in,clip]{Fig8c}
\medskip
\includegraphics[width=2.5in]{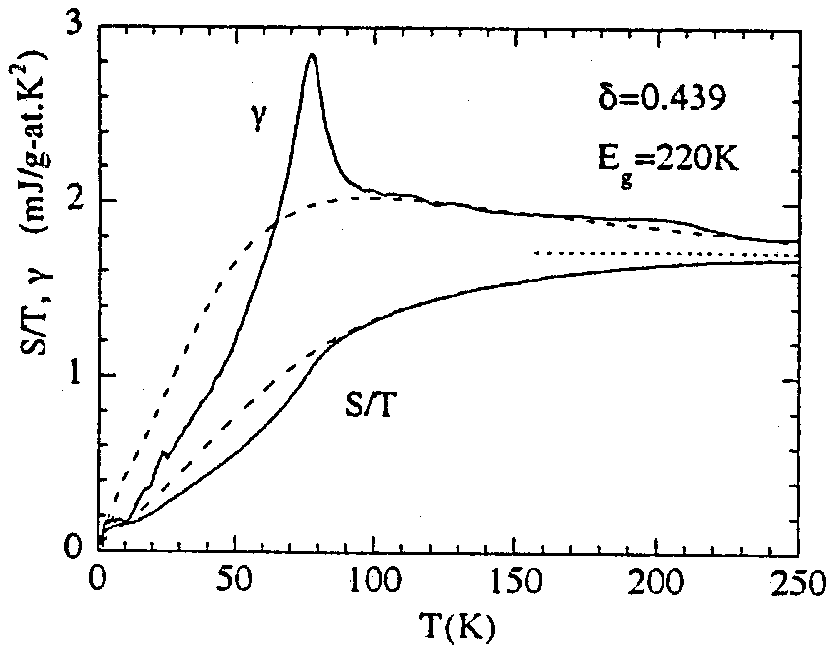}}
\caption{Extrapolated normal state (PG) and superconducting state (SC)
  contributions to SIN tunneling and thermodynamics (left), as well as
  comparison with experiments (right) on tunneling for Bi2212 from Renner
  \textit{et al.}\protect\cite{Renner} and on specific heat for
  Y$_{0.8}$Ca$_{0.2}$Ba$_2$Cu$_3$O$_{7-\delta}$ from Loram
  \textit{et al.}\protect\cite{Loram98} The theoretical SIN curve is
  calculated for $T = T_c/2$, while the experimental curves are measured
  outside (dashed) and inside (solid) a vortex core. }
\label{Renner_Loram}
\end{figure*}

In Fig.~\ref{Renner_Loram}, we plot our results for the SIN tunneling
characteristics, $dI/dV$, and the computed entropy and specific heat in
a pseudogap superconductor, and compare with
experiment.\cite{Loram98,Renner,note4} To obtain the extrapolated normal
state (called PG) in $dI/dV$, we set the superconducting order parameter
$\Delta_{sc}$ to zero, but maintain the total excitation gap to be same
as in a phase coherent, superconducting state --- with non-zero
$\Delta_{sc}$ (called SC).  This procedure presumes that when the
condensate is absent, the pseudogap $\Delta_{pg}$ must correspond to the
full excitation gap. Thus it should reflect the pairs which would
otherwise be condensed.  The characteristic behavior of $dI/dV$ measured
in an SIN configuration is presented as a comparison between theory
(left) and experiment (right) in the upper panels of
Fig.~\ref{Renner_Loram}.  Here the experimental curves are taken from
Ref.~\onlinecite{Renner}, measured for underdoped Bi2212 inside (PG) and
outside (SC) a vortex core, respectively.  The non-Fermi liquid nature
of the extrapolated normal state can be clearly seen. In a similar
fashion, we show the comparisons for the extrapolated entropy and
specific heat between our theory and experiments of
Ref.~\onlinecite{Loram98} for
Y$_{0.8}$Ca$_{0.2}$Ba$_2$Cu$_3$O$_{7-\delta}$. Here $\gamma = C_v/T$.
The agreement between the theoretically computed curves and
experimentally deduced curves provides reasonable support for the
present theoretical picture.

\subsection{Discussion}

In this section we re-visit some of the issues raised in this paper and
in experiments on $C_v$, tunneling, vortex core and related
spectroscopies. In contrast to what has been presented up until now,
here we are more qualitative and, in some instances, more speculative.

\textit{Intrinsic tunneling} experiments:\cite{Krasnov} Considerable
attention has been directed towards intrinsic tunneling experiments (in
stacked layers), not only because they yield different results from
STM\cite{Renner_Tunneling,Davis} and from point contact/break
junction\cite{JohnZ} experiments, but also because they sometimes reveal
an unexpected sharp feature (or second peak in $dI/dV$) presumably
associated with superconductivity.  This peak occurs in addition to a
broader excitation gap feature (which is more like that found in single
junction experiments) and it vanishes for $T \ge T_c$.  There is, as
yet, no complete convergence between different intrinsic Josephson
junctions (IJJ) tunneling experiments.  On overdoped samples Suzuki and
co-workers\cite{Suzuki} have found anomalously sharp and large amplitude
(second) peaks whose presence correlates with long range order, while,
by contrast, Latyshev\cite{Latyshev} and co-workers find only a single
maximum in $dI/dV$ below $T_c$, which rather smoothly evolves into the
normal state peak.  However, for underdoped samples, Krasnov and
co-workers\cite{Krasnov} report two maxima below $T_c$, with a much less
pronounced sharp feature than in Ref.~\onlinecite{Suzuki}, (although the
latter appears to be descended from the more anomalous features reported
earlier by Suzuki \textit{et al}).  Earlier work\cite{Yurgens} by some
of his same co-authors found only a single gap feature, as was
consistent with single junction data from other
groups.\cite{JohnZ,Renner}

There is in all these experiments the possibility that the so-called
superconducting peak is an artifact of self-heating or other
non-equilibrium effects, which would be present when there is a non-zero
critical current.  Its absence in single junction experiments would,
otherwise, be difficult to explain.  These latter experiments correlate
well with ARPES. Moreover, they also correlate with inferences from
$C_v$ and other bulk data.\cite{Tallon,Timusk}  We have no simple
explanation for the sharp feature in tunneling.  Within our approach
there is a single excitation gap, $\Delta$ above as well as below $T_c$.
Although the critical current\cite{Chen2} $I_c$ reflects the order
parameter $\Delta_{sc}$, which vanishes at $T_c$, this order parameter
contribution is not expected to show up in quasiparticle tunneling as a
second gap feature.  Of these IJJ experiments, the data which seems not
incompatible with our picture is that of Yurgens \textit{et
  al.},\cite{Yurgens} and possibly Latyshev \textit{et
  al.};\cite{Latyshev} the latter authors, nevertheless, find much
sharper maxima in $dI/dV$ than we would have.

\textit{Quantum critical points}: Both theorists\cite{Laughlin} and
experi\-mentalists\cite{Valla,Loram98} have recently turned their
attention to quantum critical phase transitions.  Moreover, these
phenomena are assumed to be related to pseudogap effects. Indeed, Loram
and co-workers\cite{Loram98} presume that $\Delta_{pg}$ (which, in their
approach, is taken to be temperature independent) is proportional to
$T^*$. Then at some critical doping concentration ($x \approx 0.19$),
$\Delta_{pg}$ appears to vanish and they infer that $ T^* \rightarrow
0$.  By contrast, we find $\Delta_{pg}$ is more closely associated with
$(T^* - T_c)$, which does not lead to a zero temperature phase
transition, even when $\Delta_{pg}$ vanishes.

At low $x$, on the other hand, there may be something more dramatic like
a first order or quantum critical phase transition going on --- at the
superconductor-insulator boundary.  Here the excitation gap is maximum
on one side of the boundary and yet the superconducting order parameter
disappears on the other. [In the present picture this disappearance was
found to arise from the localization of $d$-wave pairs\cite{Chen1}]. At
low $x$, this superconductor-insulator transition also appears in the
presence of a magnetic field\cite{Boebinger} when the field is large
enough to drive the system into the normal phase. It also appears with
impurity pair breaking.\cite{Zhao} All three of these experiments may be
interpreted as suggesting that the fermionic excitation gap $\Delta$
survives in the presence of pair breaking (by large fields or
impurities), or low hole concentrations --- thereby leading to an
insulating fermionic excitation spectrum. Some confirmation of this
conjecture comes from NMR experiments\cite{Pennington} which seem to
imply that $\Delta$ does not vary with $H$, once the pseudogap is well
established.  And the STM measurements,\cite{Renner} in general, as well
as inside a vortex core seem consistent with the observation that
$\Delta$ is only weakly $H$ dependent. One cannot, of course, ignore
Mott insulating effects at this superconductor-insulator boundary as
well.  However, whatever physical mechanism is dominant, the fact that
the superconductor-insulator transition is a robust feature associated
with the disappearance of superconductivity, whether by pair-breaking or
by doping, needs to be addressed.

\textit{Incoherent contributions to the spectral function in
  photoemission}:

Recent photoemission experiments\cite{Campuzano,Shennew} indicate that
the relative weight of the coherent contribution to the spectral
function decreases rapidly with decreasing $x$.  Ignoring the incoherent
term $\Sigma_0$ in the self energy (as we have here) will necessarily
affect any quantitative inferences about the systematic $x$ dependence
of the spectral function and, thereby, of $C_v$ or the related
condensation energy. This term can only be put in by hand in the present
approach; it arises here from diagrams other than the particle-particle
terms which give rise to the superconductivity and pseudogap.  Moreover,
the measured systematic $x$ dependence of the coherent spectral weight,
which has been inferred from photoemission,\cite{Campuzano,Shennew} is
likely to be consistent with the inferences based on
thermodynamics\cite{Loram98} for the $x$ dependence of the entropy $S$
and related condensation energy. However, when the contribution from
$\Sigma_0$ is sizeable, relative to the coherent terms, one cannot
include it (by hand) without self-consistently also re-solving for the
chemical potential $\mu$ and, hence, also for $\Delta$ and $T_c$, etc.
This extensive numerical program would take us too far afield to
implement here.

In addition, this $\Sigma_0$ contribution is needed to arrive at the
well known\cite{Norman} dip-hump features of photoemission.  When this
term is artificially added, we are able to obtain this latter structure
which will scale with the excitation gap $\Delta$.  Indeed, a dependence
on $\Delta$ is plausible since we presume that $\Sigma_0$ is associated
with various (electron-hole) polarizabilities, which in the
superconducting and pseudogap states reflect the non-vanishing
excitation gap.  It is, however, essential (to obtain dip-hump features)
that the imaginary part of $\Sigma_0$ turns on rather abruptly at
frequencies somewhere between $\Delta$ and $ 2 \Delta$.  Indeed, the
step function model introduced in Ref.~\onlinecite{Norman3} seems to
accomplish this quite well, but we know of no simple microscopic
mechanism which yields this rapid frequency onset.

\section{Conclusions}

This paper has raised some issues which have a number of important
ramifications. We suggest that specific heat and vortex core experiments
have provided strong evidence that the normal state underlying the
superconducting phase is not a Fermi liquid. Thus BCS theory cannot be
applied to the underdoped cuprates, without some modification.  Since we
have, as yet, very little alternative to BCS, it is natural first to try
to extend it slightly.  We believe the simplest and most benign
modification is to adopt the Leggett extension of the ground state
wavefunction, Eq.~(\ref{eq:1}), (which is applicable to weak and strong
coupling $g$) and extend his scheme to finite $T$. Within this approach
we were able in the past to perform a number of concrete
calculations,\cite{Kosztin1,Kosztin2,Chen1,Chen2,Chen3} and, in the
present paper, explore the behavior somewhat above and below $T_c$ of
the specific heat, $C_v$, and quasiparticle tunneling characteristics,
$dI/dV$.  This present study led us naturally to analyze the nature of
superconducting phase coherence in the presence of a pseudogap.

Our study of $C_v$ and $dI/dV$ is based on a Green's function decoupling
scheme chosen to be consistent with Eq.~(\ref{eq:1}). The spectral
functions which enter these two physical quantities depend, in turn, on
the self energy which is the sum of Eqs.~(\ref{SigmaPG_Model_Eq}) and
(\ref{SigmaSC}), where throughout we have ignored the incoherent term
$\Sigma_0$, which enters through diagrams other than those responsible
for the superconducting and pseudo-gaps.  It should be noted that this
form for $\Sigma$ is different from that introduced phenomenologically
by Franz and Millis\cite{Franz} and by Norman.\cite{Norman3}  One of
these groups,\cite{Franz} in particular, emphasized the effects of the
temperature dependent scattering rate.  Here, by contrast, we emphasize
the effects of long range phase coherence which sets in at $T_c$. 

Because of the breakdown of BCS theory, in a superconductor with a
pseudogap, the standard simplifications, such as Landau-Ginzburg
expansions and Bogoliubov-de Gennes approaches\cite{Tesanovic} are not
expected to hold, at least without some modifications.  For the BCS
case, the expansion in terms of a small order parameter which is
identical to the excitation gap at $T_c$ is possible.  However, when the
order parameter and the excitation gap are distinct as in the pseudogap
case, there is no straightforward way to expand the free energy in terms
of a small order parameter and to reflect the existence of a well
established excitation gap simultaneously.

It should be emphasized, finally, that, in the non-BCS superconductor,
there is an important distinction between $T_c$ and the zero temperature
excitation gap $\Delta(0)$.  In the strong coupling, but still fermionic
regime, as the pseudogap increases, $T_c$ is
suppressed.\cite{Maly1,Maly2,Chen1} This observation allows us to respond to
the widely repeated criticism of this ``preformed" pair approach:
namely that\cite{Pennington} ``the gap is not closely tied to the onset
of superconductivity" --- as inferred from, say, the lack of magnetic
field dependence in the former.  Here we claim that in contrast to BCS
theory, the excitation gap $\Delta$ is expected to be robust with
respect to standard pair-breaking perturbations (such as magnetic fields
and impurity scattering), while the order parameter, $\Delta_{sc}$, is
not.  As emphasized throughout this paper, the distinction between these
two quantities is an essential component of the ``preformed" pair
approach.

This work was supported by the NSF-MRSEC, No.~DMR-9808595 (QC and KL),
and by the University of Illinois (IK).  We thank A.~V. Balatsky, K.~E.
Gray, Ying-Jer Kao, J.~W. Loram, and J.~F. Zasadzinski for helpful
discussions.

%\bibliographystyle{prsty} 
%\bibliography{QChen}

\begin{thebibliography}{10}
\bibitem{note1}
We make a distinction here between tunneling and ARPES in the presence of an
  isotropic tunneling matrix element. For more directional tunneling, the
  difference between ARPES and tunneling experiments may be less apparent.
\bibitem{gamma_note}
For the $d$-wave case, we have also studied the effects associated with
  introducing a {\bf k}-dependence in $\gamma$, and found that these effects
  are negligible.
\bibitem{newfootnote}
Here by $\mbox{d}N(\omega)/\mbox{d}T$, we mean the Fermi function weighted
  average over a broad range of frequency. See the third term in
  Eq.~\ref{Cv_Eq}.
\bibitem{note2}
It should be emphasized that the temperature dependence of $C_v$ at $T\gtrsim
  T_c$ is dominated by $\mbox{d}N(\omega)/\mbox{d}T$, instead of $N(\omega)$
  itself. In fact the contributions of these two functions are in opposite
  directions as a function of temperature. This observation implies that one
  cannot extract the size of the excitation gap from the specific heat without
  taking proper account of the $T$ dependence of the density of states. It
  should also be noted that the curvature above, but in the vicinity of $T_c$
  also reflects the subtle feature at $T_c$ which appears in the excitation
  gap.
\bibitem{note3}
For this choice of $t^\prime$, the van Hove singularities appear inside the
  filled lower half band. Because these effects were not particularly relevant
  in our previous studies, this complication was ignored previously.
\bibitem{Power}
Since the pseudogap equation (\ref{eq:gap3}) indicates $\Delta_{pg}^2\propto
  T^{3/2}$ at low $T$, and $\Delta_{pg}^2$ is roughly linear in $T$ at
  $T\lesssim T_c$, this choice of $\gamma\propto T^3$ below $T_c$ will give
  rise to a temperature dependence of the effective quasiparticle peak width
  $\gamma^\prime \propto T^\alpha$ with $\alpha=4\sim 4.5$ for $T\leq T_c$, via
  Eq.~(\ref{Gamma'_Eq}). This is consistent with the experimentally observed
  $T^4 \sim T^5$ dependence of the quasiparticle scattering rate.\cite{Hardy2}
  Above $T_c$, the $T$-dependence of $\Delta_{pg}$ is weak, so that
  $\gamma^\prime \propto T$, in agreement with ARPES data.\cite{Valla}.
\bibitem{note4}
In what follows we ignore the contribution from bound states to the core
  tunneling spectroscopy. Understanding these bound states will require a
  generalization of Bogoliubov-de~Gennes theory for our extended BCS approach.
\end{thebibliography}

\end{document}